\documentclass[showpacs, twocolumn]{revtex4-1}
\usepackage{graphicx}
\usepackage{amsmath}

\begin{document}

\title{ Anderson localization of quantum droplets in disordered potentials}

\author{ Zohra Mehri$^{1}$ and Abdel\^{a}ali Boudjem\^{a}a$^{2}$}

\affiliation{$^1$ Department of Physics, Faculty of  Sciences and Technology, Ahmed Zabana University of Relizane, Bourmadia, P.O. Box 48000,Relizane, Algeria.\\
$^2$ Department of Physics, Faculty of Exact Sciences and Informatics, Hassiba Benbouali University of Chlef P.O. Box 151, 02000, Ouled Fares, Chlef, Algeria}

\email{a.boudjemaa@univ-chlef.dz}

\begin{abstract} 

We study  Anderson localization of a one-dimensional quantum droplet in a speckle-like potential employing the generalized Gross-Pitaevskii equation.
We compute the droplet width, density profiles, diffusion exponent and coefficient, and the localization length for both small and large droplets.
Interesting classes of anomalous diffusions are obtained in transport dynamics ranging from superdiffusion to subdiffusion for a strong disorder strength.
We find that above a certain critical disorder strength the droplet exhibits a transition to Anderson localization. 
Our results can be redibly probed with recent experiments.

\end{abstract}
\maketitle

\section{Introduction}

The exploration of Anderson localization (AL) of expanding  Bose-Einstein condensates (BECs) in random media has gained traction, 
both from the theoretical and experimental point of views (see for review \cite{Asp,Modu}).
AL can be understood as the effect of interference between multiple-scattering paths  where particles remain localized with exponentially decaying density \cite{Anderson}. 
During the last years,  AL of matter waves exposed to one-dimensional (1D)  disordered potentials has been widely investigated, both in the discrete \cite{Roat} 
and in the continuum \cite{Billy} setups. 
The expansion of BECs in 2D \cite{Wit, Naj} and in 3D (see e.g. \cite{Kuhn,Skip,Kond, Jend, Semeg,Yed1,Yed2}) random environments 
has also attracted considerable attention in a quest for observing AL.

One of the remarkable recent achievements in the field of cold atoms is the prediction and the realization of quantum droplets in both Bose mixtures \cite{Petrov, Cab, Sem} 
and dipolar BECs \cite{ Pfau,Pfau1, Chom}.
The formation of such exotic structures relies on the balance between  attractive mean-field forces, and repulsive beyond mean-field interactions
provided by the Lee-Huang-Yang (LHY) quantum fluctuations \cite{LHY}.
Recent research has demonstrated that the presence of disorder (depending on its strength) can modify the shape of these droplets and  destabilize them, 
causing them to split into several mini droplets \cite{Sahu,Abbas1,Abbas2,Abbas3}. 
Furthermore, AL of Bogoliubov quasiparticles in disordered quantum droplets has been deeply studied in our most recent work \cite{Mehri}.

Despite significant progress in understanding the properties of disordered atomic ultracold  gases, 
AL of quantum liquids spreading in a disordered potential remains rarely  explored in the literature. 
The aim of this work is then to investigate the dynamic evolution of an initially confined interacting 1D quantum droplet in a symmetric Bose-Bose mixture  
and subjected to a disorder potential produced by an optical speckle-like pattern.
Quantum droplets are self-bound and ultradilute states of matter allowing the exploration of AL in a very controlled manner compared to BECs.
In a 1D configuration, self-bound droplets appear in two different scenarios: small droplets which have  Gaussian-like shape for small number of particles  and 
large droplets with a flat-top plateau for a large number of atoms \cite{Petrov1, Astra}.

Here, we calculate standard observables such as the droplet width, density profiles, diffusion exponent and coefficient, and the localization length 
for both small and large droplets by numerically  solving the underlying time-dependent generalized Gross-Pitaevskii equation (TDGGPE).
In the disorder-free expansion, the droplet maintains its self-bound character and remains almost stable during the subsequent evolution.
However, our results suggest that the inclusion of the disorder refrains the self-bounding process by providing an additional energy to atoms 
causing, depending on the disorder strength, the fast expansion and eventually the emergence of an anomalous diffusion.
The resolution of different transport regimes simultaneously enables us to extract the expected diffusion exponent and coefficient. 
We find that above a certain critical disorder strength, the droplet exhibits signatures for a transition to AL. 
We emphasize that the localization length and time depend on the size of the droplet.

The remainder of the paper is organized as follows.
In Sec.~\ref{Mod} we provide a brief description of the TDGGPE model and the speckle disorder potential.
Section~\ref{NR} deals with numerical results obtained from the 1D TDGGPE and discusses the AL. 
We analyze the droplet width, density profiles, diffusion exponent and coefficient, and the localization length.
Finally, we draw our conclusions in Sec.~\ref{conc}.

\section{Model} \label{Mod}

We  consider a 1D quantum droplet in a symmetric Bose-Bose mixture confined in a harmonic potential $V({z})$ 
with trap frequency $\omega$, and subjected to an external random potential $U({z})$.
The dynamics of the  system is governed by the TDGGPE:
\begin{equation} \label{GGPE}
i\hbar \frac{\partial \psi}{\partial t} =\bigg[-\frac{\hbar^2}{2m} \frac{\partial^2}{\partial z^2}+V({z})+ U({z})+\delta g|\psi|^2 - \frac{2m g^{3/2}}{\pi \hbar} |\psi|\bigg] \psi,
\end{equation}
where $\psi\left( { z},t\right)$ is the wavefunction, and the parameter $\delta g$ is related
to the intraspecies, $g$,  and interspecies, $g_{12}$, coupling constants as $ \delta g = g_{12}+g$.
The disorder potantial which is assumed to have vanishing ensemble averages $\langle U(z)\rangle=0$
and a finite correlation of the form $\langle U(z) U(z')\rangle=R (z-z')$.
The speckle pattern is created by deflecting a laser beam through a rough plate. 
However, theoretically the speckle potential can be generated using different methods (see e.g. Refs. \cite{Asp,Modu,Pilat,Adh, Abdu} and references therein).
In our case, we model the potential $U(z)$ by a set of $S$ identical Gaussian spikes randomly distributed along the $z$-axis \cite{Asp,Adh}:
\begin{equation} \label{Spk}
	U (z)=U_{0}\displaystyle\sum_{i=1}^{S} U_{\text{dis}}( z- z_{i}),  
\end{equation} 
where $U_{0}$ is the strength of the disorder, and $z_{i}$ are the uncorrelated random positions, $S$ is the number of impurities, 
and $U_{\text{dis}}$ is a real valued-function of correlation length $\sigma$ which measures the spread of the correlations, and has Gaussian shaped impurities 
$U_{\text{dis}}(z)=e^{-z^2/\sigma^2}/(\sigma \pi^{1/2})$. These spikes can overlap creating a speckle-like pattern.

It is convenient to express Eq.~(\ref{GGPE}) in terms of dimensionless quantities: $z=z_0 z$, $ t= t_0 t$, $\omega= \omega_0 \,\omega$, and $\psi=\psi_0 \psi$,
where $ z_0=\pi \hbar^2 \sqrt{\delta g/2} /(m g^{3/2})$,  $t_0=\pi^2 \hbar^3 \delta g/(2m g^3)$, $\omega_0= t_0^{-1}$, and 
$ \psi _0 =\sqrt{2m} g^{3/2}/(\pi \hbar \delta g )$. This yields
\begin{equation} \label{GGPE1}
i \frac{\partial \psi}{\partial t}=\left(-\frac{1}{2} \frac{\partial^2}{\partial z^2} +V(z)+ U(z) +| \psi|^2-|\psi| \right) \psi.
\end{equation}
In the absence of the disorder and trapping potentials, the ground-state solution of the TDGGPE takes the explicit form \cite{Petrov1}:
\begin{equation} \label{EqSol}
\psi(z,t)=-\frac{3\mu \exp(-i\mu t)}{1+\sqrt{1+9\mu/2} \cosh(\sqrt{-2\mu}z)},
\end{equation}
where $\mu$ is the chemical potential related to the norm as: 
\begin{equation}
N=\frac{4}{3}\bigg[\ln\bigg(\frac{\sqrt{-9\mu/2}+1}{\sqrt{9\mu/2+1}}-\sqrt{-9\mu/2}\bigg)\bigg],
\end{equation}
which is proportional to the number of atoms in the droplet.
A broad flat-top (large) droplet corresponds to $\mu=-2/9$ and a constant inner density  $n=4/9$ \cite{Astra}.

\section{Numerical results} \label{NR}

In our simulations, initially the droplet is prepared in the ground-state of Eq.~(\ref{GGPE1}). At $t = 0$, the harmonic potential is switched off  so the droplet evolves in the disorder potential.
For computing this evolution we use the split-step Fourier method (SSFM).
The trap frequency is assumed to be low to ensure the robustness of the flat-top density of large droplets, and 
the chemical potential $\mu$ must be tuned close to the  particle emission threshold $-2/9$ in order to obtain the desired value of $N$.
The system consists of a spatial grid with $L=200$ sites with spatial and time steps $\Delta z=1$, $\Delta t =0.5$, respectively,
and the simulation is averaged over $100$ different realizations of disorder to obtain statistically meaningful results.
Time evolution is implemented using the SSFM, which alternates between applying the nonlinear interaction including the LHY correction, disorder potential in real space 
and the kinetic term in Fourier space. 
The SSFM algorithm is highly efficient for solving the TDGGPE  and provides accurate and stable solutions for quantum droplets \cite{Abbas2}.

\begin{figure}[hh]
		\includegraphics[scale=0.8]{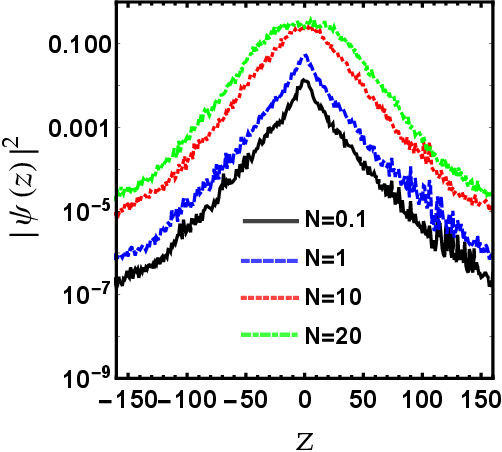}
	\caption{Stationary density profiles of the droplet after $t=200$ subjected to a speckle-like potential for different values of  $N$.
Parameters are: $U_0 = 3$ and $\sigma =0.5$.
Here we set $\mu=-0.1493$, $\mu=-0.1954$, $\mu =-0.222222$, and $\mu = -0.2222222222222$
 to obtain $N=0.1$, $N=1$, $N=10$, and $N = 20$, respectively.}
\label{dprof}
\end{figure}

In Fig.~\ref{dprof} we present the stationary density of the droplet in the disorder potential of Eq.~(\ref{Spk}) for different values of $N$.
We can observe that for the smallest value, $N= 0.1$ (small droplet), the density decays exponentially in the tails and develops a sharp peak in the center 
which is very reminiscent of AL observed in a disordered BEC \cite{Asp,Modu}.
However, for large $N$ (i.e. large droplet), the density decays exponentially and features a chapeau in the center of the sample.
In the tails of the stationary profile, the droplet is very dilute, and the kinetic energy exceeds both mean-field and beyond mean-field energies.

\begin{figure}[hh]
		\includegraphics[scale=0.45]{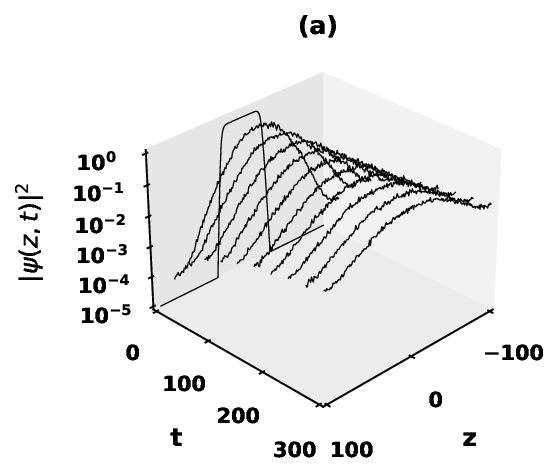}
		\includegraphics[scale=0.45]{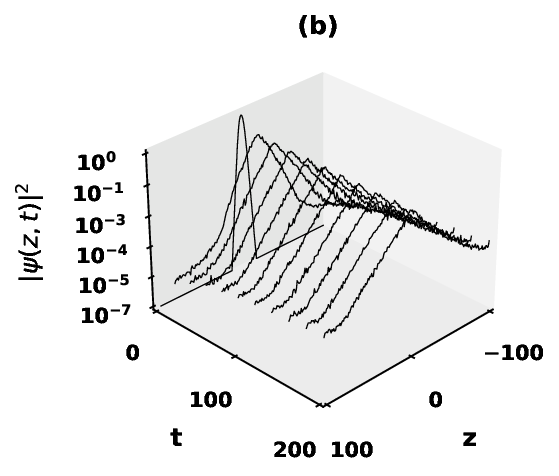}
		\caption{Time evolution of quantum droplets (on a log scale) in the speckle-like potential for $U_0=2$ and $\sigma=0.5$.
 (a) Large droplet with $N=20$. (b) Small droplet with $N=1$.}
\label{Dplot}
	\end{figure}

\begin{figure}[hh]
\includegraphics[scale=0.45]{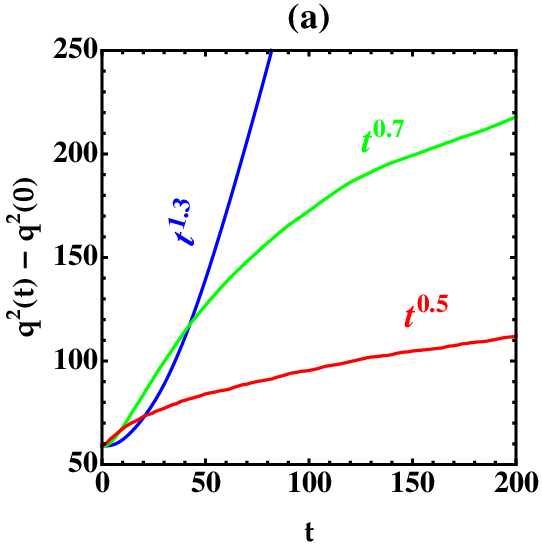}
 \includegraphics[scale=0.45]{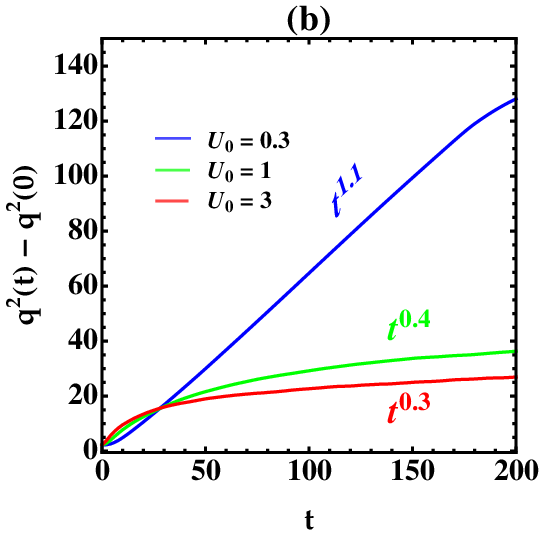}
	\caption{ (a) Time evolution of the width, $q^2(t)-q^2(0)$, of a large droplet, $N=20$, for several values of $U_0$ with $\sigma =0.5$.
(b) The same but for a small droplet, $N=1$.}
\label{size}
\end{figure}

\begin{figure}[hh]
	\includegraphics[scale=0.8] {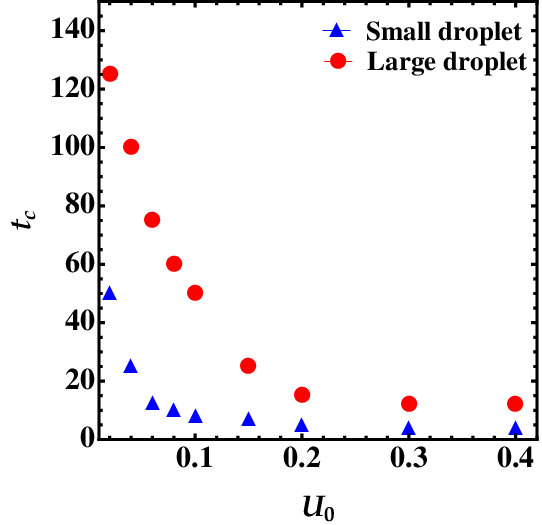}
		\caption{ Critical time  as a function of the strength of disorder  for small  (triangles) and large  (circles) droplets.}
\label{crtTime}
\end{figure}

The dynamics of the density profile for both large and small droplets released from a harmonic potential  in a 1D  speckle-like disorder potential 
is depicted in Fig.~\ref{Dplot}.
Experimentally, the dynamics of atoms can be captured by evaluating absorption-image technique \cite{Modu,Barb}.


It is apparent from Fig.~\ref{Dplot} (a)  that when released into a disorder potential, the large droplet 
maintains its flat top shape up to a certain critical time (see below), then it undergoes genuine AL due to the presence of disorder.
Practically, the same behavior holds for small droplet, where  the cloud evolves to an exponential profile on the tails during the evolution as seen in Fig.~\ref{Dplot} (b).
The most striking observation in these numerical data is that during the evolution, the quantum liquid remains localized in the center of the sample and oscillates around its equilibrium value.
On the other hand,  the distribution of both structures exhibits a non-exponential behavior on the tails revealing the suppression of AL at longer times.
In such a regime, one may expect a destruction of a long-range coherence due to the interplay of quantum fluctuations and randomness,
leading to prevent AL and thus classical (anomalous) diffusion might be started.

\begin{figure}[hh]
	\includegraphics[scale=0.43]{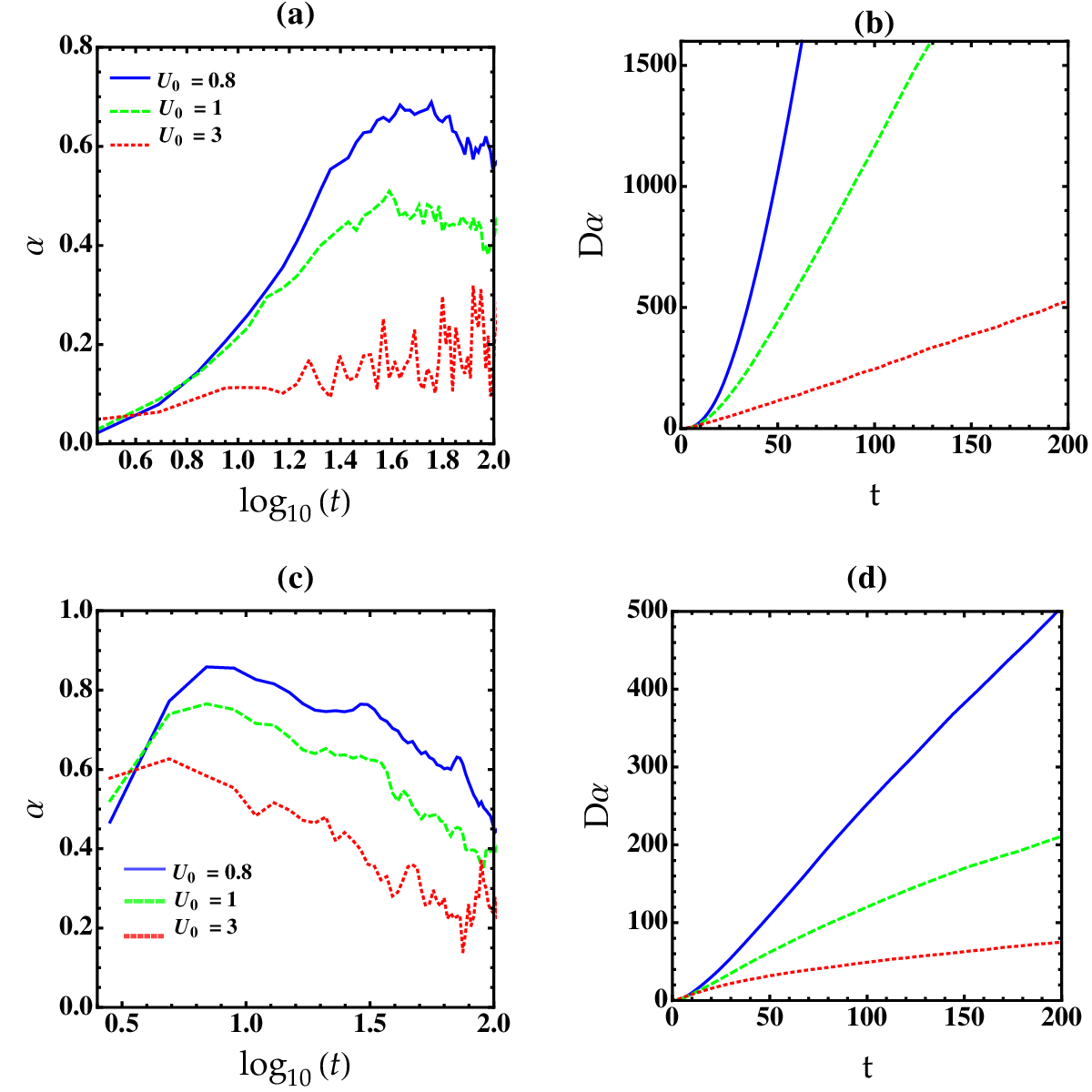}
	\caption{ (a) The exponent $\alpha $ as a function of time taken from Fig.~\ref{size} for several values of  $U_0$ with $\sigma =0.5$. 
(b) Time dependence of the diffusive cofficient, $D_{\alpha}$, for several values of $U_0$ with $\sigma =0.5$.
(c) and (d) are the same as (a) and (b) but for a small droplet.}
\label{expcoef}
\end{figure}

\begin{figure}[hh]
	\includegraphics[scale=0.45]{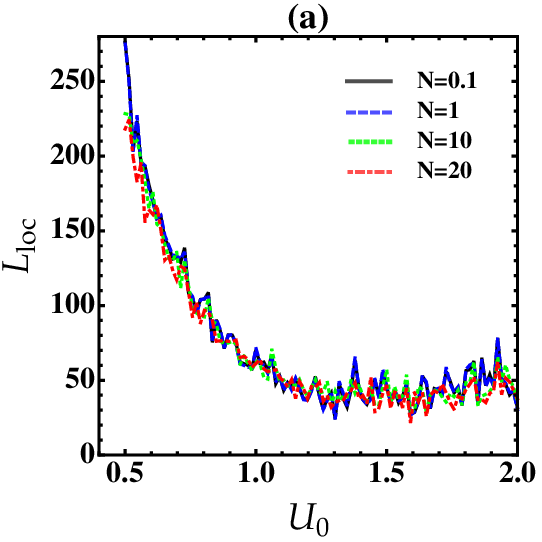}
	\includegraphics[scale=0.45]{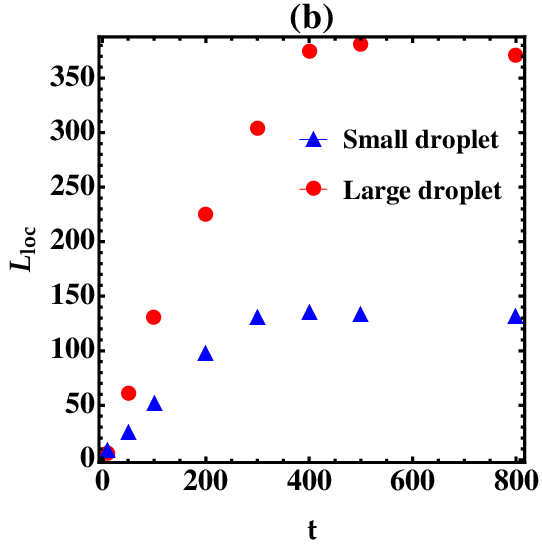}
	\caption{ (a) Localization length, $L_{\text{loc}}$, versus the disorder strength for different values of $N$.
(b) Time evolution of the localization length for large (circle) and small (triangles) droplet and BEC (squares).
Parameters are: $U_0=2$ and $\sigma =0.5$.}
\label{LLoc}
\end{figure}

To gain further insight into these, we present in Fig.~\ref{size} the typical dynamics of the droplet width, $q^2(t)= \int |\psi(z,t)|^2 z^2 dz$, for different values of disorder strength.
We observe two transient transport regimes.
Initially, the size is constant up to a certain critical time, $t_c$, meaning that the droplet reaches its equilibrium state. 
The plateau observed in the evolution of the equilibrium width of the droplet strongly depends on the disorder strength  (see Fig.~\ref{crtTime}).
For instance for a weak disorder $U_0=0.05$, the critical time  $t_c \simeq 100$ while it decreases as the disorder strength increases until it saturates to $t_c =15$ for $U_0 \geq 0.2$.
After this transient i.e. at $t>t_c$, the droplet becomes weakly self-bound and evaporates eventually, thus it expands as a gas (i.e. exhibits a power-law growth)  as forseen above.

To model the expansion, we fit the obtained evolution with the solution of a generalized diffusion equation:
\begin{equation} \label{ExpLaw}
q^2(t)= q^2(0) + 2D_{\alpha} t^\alpha,
\end{equation}
where $q(0)$ is the initial width, $\alpha$ is the diffusion exponent, and $D_{\alpha}$ is a generalized diffusion coefficient.
For $\alpha=1$, the system undergoes normal diffusion, subdiffusion for $\alpha< 1$, and superdiffusion for $1< \alpha <2$.
Other special cases, for $\alpha=2$, the system spreads ballistically, and $\alpha> 2$ is called hyperballistic describing intriguing expansions.
The diffusion law (\ref{ExpLaw}) can correctly model the overall behavior from short times to the asymptotic regime.

Interesting classes of anomalous diffusions are found in transport dynamics of droplets (see Fig.~\ref{size}).
(i) For the disorder strength, $U_0=0.3-1$, we extract from the fit  $\alpha \approx 1.3 -0.7$, indicating that the dynamics of both droplets is almost diffusive.
(ii) For stronger disorder, $U_0=3$, the width increases according to Eq.~(\ref{ExpLaw}) subdiffusively ($\alpha \approx 0.3$), 
which could be attributed to signatures of AL.

In order to quantitatively analyze the change of the droplet spreading dynamics seen in Fig.~\ref{size},
we calculate the time-dependent  exponents $\alpha (t) =d\log_{10}\Delta q(t)/(d\log_{10}t)$ and 
the time-dependent anomalous diffusion coefficient as $D_{\alpha} (t)= \langle q_i^2(t)-q_i^2(0)/(2t^{\alpha}) \rangle$ \cite{Barb}.
The results for both large and small droplets are reported in Fig.~\ref{expcoef}, where one observes a clear decrease of both $\alpha$ and $D_{\alpha}$ with disorder strength over time.
At short times, $\alpha$ increases and then saturates (weakly varying) at longer times.
For instance, at $t\gtrsim 150$, it attains a saturation value around $\alpha \simeq 0.2$ for a large droplet while $\alpha \simeq 0.25$ for a small droplet 
in the case of a strong disorder which is smaller than previously observed for discrete models \cite{Modu}  giving rise to the expansion of wavepacket into its localized state.
The increase of $\alpha$ and $D_{\alpha}$ notably for weak disorder reflects that the system acquires a higher kinetic energy,
revealing the persistence of diffusive wave propagation in the  speckle-like potential.

Another interesting observable to unveil signatures for a transition to AL above a critical disorder strength is the localization length, $L_{\text{loc}}$ which is the inverse of the Lyapunov exponent $\gamma=1/L_{\text{loc}}$. 
It can be extracted from an exponential fit of the tails of the density profile as $|\psi(z)|^2\sim \exp(-2|z|/L_{\text{loc}})$ \cite{Laurent}.
In Fig.~\ref{LLoc} we present the localization length for different values of $N$.
As shown in Fig.~\ref{LLoc} (a), $L_{\text{loc}}$ is decreasing with increasing $U_0$ for both large and small droplets leading to more localized behaviors. 
We see from Fig.~\ref{LLoc} (b) that $L_{\text{loc}}$ increases linearly over time then it saturates at a later time $t\geq t_{\text{loc}}$ (localization time) towards a constant value. 
Such a saturation could be a fingerprint of the AL phenomenon.
It is also clearly visible from the figure that $L_{\text{loc}}$ associated with a small droplet is smaller than that of a large droplet implying that the former supports higher localization 
which weakens as we increase the size of droplet, $N$. 
Another remarkable feature is that the small droplet localizes ($t_{\text{loc}} \simeq 300$) faster than the large droplet ($t_{\text{loc}} \simeq 400$).

For example, if one considers a typical set of experimental parameters for ${}^{39}$K atoms with intra- and interspecies scattering lengths $a = 69.99 a_0$,
$ a_[12]= - 53.37 a_0$  \cite{Sem}  where $a_0$ being the Bohr radius, and fix the disorder correlation and strength as $\sigma= 0.5 \,\mu$m and $U_0=40.7$ kHz,  
the approximately predicted values of the localization lengths after $t=0.6$\,s are : $L_{\text{loc}}=866.2\, \mu$m for a large droplet,  and $L_{\text{loc}}=770.5\, \mu$m for a small droplet.

\section{Conclusions}\label{conc}
              
In summary, we investigated the dynamic evolution of a 1D quantum droplet in the presence of a random optical speckle-like potential
by numerically solving the TDGGPE, to observe unambiguously AL in both small and large droplets.
We analyzed the different parameters which fulfil the observation of AL, such as the density profiles, the droplet width, the diffusion exponent and coefficient, and the localization length.
Our findings reveal that depending on the disorder strength and on the size of the droplet, 
the system exhibits intriguing board of anomalous diffusion, ranging from subdiffusion to superdiffusion.
Additionally, we find that localization length and time vary with respect of the droplet size.
The emerging results not only provide insights into how disorder influences the localization of quantum droplets, 
but offer also a deeper understanding of peculier quantum systems in random environments.
An interesting future prospect include the exploration of the transport and localization in quantum droplets under the influence of  time-dependent disorder potentials.

\section*{Acknowledgments}
We are grateful to Dmitry Petrov and Axel Pelster for valuable discussions.
We would like to thank the LPTMS-Paris and the Department of Physics-University of Kaiserslautern, for a visit, during which a part of this work was conceived.

\end{document}